\documentclass[letterpaper]{article} 
\usepackage{aaai2026}  
\usepackage{times}  
\usepackage{helvet}  
\usepackage{courier}  
\usepackage[hyphens]{url}  
\usepackage{graphicx} 
\usepackage{natbib}  
\usepackage{caption} 
\usepackage{algorithm}
\usepackage{algorithmic}
\usepackage{booktabs}
\usepackage{tcolorbox}
\usepackage{subcaption}

\usepackage{newfloat}
\usepackage{listings}
\DeclareCaptionStyle{ruled}{labelfont=normalfont,labelsep=colon,strut=off} 
\floatstyle{ruled}
\newfloat{listing}{tb}{lst}{}
\floatname{listing}{Listing}

\newcommand{\preprintnotice}[1]{%
    \noindent
    \begin{tcolorbox}[
    colback=blue!5!white,
    colframe=blue!75!black,
    title=Preprint Notice] {#1}
    \end{tcolorbox}
}
\title{Moral Missions: Sustaining Responsible Practice by Surfacing Practitioner Moral Decision-Making Procedures}
\author{
    Teanna Barrett\textsuperscript{\rm 1}, 
    B. Biira\textsuperscript{\rm 2}, 
    Jainaba Jawara\textsuperscript{\rm 3}, 
    Andrew Shaw\textsuperscript{\rm 1,4}, 
    Ziwei Dong\textsuperscript{\rm 5}, 
    Chinasa T. Okolo\textsuperscript{\rm 6}, 
    Seyi Olojo\textsuperscript{\rm 7}, 
    Keerthana Kompella\textsuperscript{\rm 1}, 
    Khadija Saho\textsuperscript{\rm 2}, 
    Amy X. Zhang\textsuperscript{\rm 1},
    Leilani Battle\textsuperscript{\rm 1}
}
\affiliations {
    \textsuperscript{\rm 1}Paul G. Allen School of Computer Science \& Engineering, University of Washington\\
    \textsuperscript{\rm 2}Information School, University of Washington\\
    \textsuperscript{\rm 3}University of Maryland College of Information\\ 
    \textsuperscript{\rm 4}Department of Computer Science, Cornell University\\ 
    \textsuperscript{\rm 5}Microsoft\\
    \textsuperscript{\rm 6}Technecultura\\
    \textsuperscript{\rm 7}School of Information, University of California Berkeley\\
    	tb23@cs.washington.edu\footnote{Corresponding author's email}
}

\begin{document}

\maketitle

\begin{abstract}
    A growing ecosystem of techniques, toolkits, and guidelines has been developed to help data scientists consider the social implications of data-driven technologies. However, prior literature highlights that even when this ecosystem of techniques is provided to professional data scientists, they still struggle to consistently adopt a responsible data science practice. We posit that the key to sustained responsible data science practice is to approach it as a ``\textit{moral mission}'': a conviction-driven technical practice that seeks to transform social conditions by any degree possible. In this paper, we present a semi-structured interview study with 15 responsible data scientists and AI practitioners to understand the moral decision-making procedures they use to articulate and actualize their moral missions. Through a phenomenological analysis of our participants' accounts, we find participants engage in embodied introspection, circumvent institutional expectations, and center relationality throughout their moral missions. We also present how our participants engage in similar processes to contend with generative AI (GenAI) in their responsible practice. We conclude by calling for subversive data science communities and identifying sociotechnical design implications to better support sustainable responsible data science practice.
\end{abstract}

\preprintnotice{A version of this work will be published in the proceedings of the 9th Annual AAAI Conference on AI, Ethics, \& Society (AIES).}

\section{Introduction}
Despite a growing ecosystem of frameworks, toolkits, and guidelines for responsible data science (RDS), practitioners struggle to meaningfully integrate these resources into everyday practice~\cite{wong2023seeing}. One obstacle to responsible practice is the deprioritization of RDS techniques within organizations, particularly amid shifting political and institutional interests that impact sustained attention to artificial intelligence (AI) ethics. Consequently, upholding and actualizing RDS is often left to individual practitioners \cite{metcalf2019owning}. Prior work has largely focused on outlining how practitioners should practice RDS; less scholarship pays attention to how practitioners make moral decisions in situ~\cite{madaio2020co-designing,madaio2022assessing}. 

Moral decision-making procedures capture the experience of engaging with a set of principles, codes, or values to guide one's actions~\cite{shipp2023making, smith2018using, preti2018developing}. From a pragmatist perspective, these processes are rarely idealized applications of right or wrong but complex assessments of the situation, available decisions, and the applicability of one’s convictions \cite{jafarinaimi2015values}. Individual moral decision-making procedures that mediate between values and technical action remain underexplored \cite{wang2024strategies, ali2023walking}. This gap is critical as responsible practice is not solely determined by the availability of tools, but by the situated judgments, constraints, and value negotiations that shape their use within organizational contexts. 

We argue that supporting RDS requires understanding moral decision-making procedures as an explicit part of RDS practice.
To this end, we focus our study of RDS practice on capturing defining decisions that are often overlooked and implicit strategies that sustain responsible practice.

We present a qualitative phenomenological study of fifteen self-identified responsible data scientists to surface the moral reasoning procedures that guide applications of responsible techniques under real-world constraints. We draw on interpretive phenomenological analysis (IPA) to investigate the lived moral reasoning that guides how practitioners select, adapt, and deploy RDS techniques. 

Our study is guided by the following research questions: 
\textbf{
\begin{enumerate}
    \item How do data scientists committed to building responsible data-driven technology identify their personal values?
    \item How do data scientists committed to building responsible data-driven technology adapt and integrate responsible data science techniques into their work?
\end{enumerate}
}

From our participants' accounts of practicing RDS, we conceptualize the phenomenon as a ``\textit{moral mission}''. We define moral missions as conviction-driven technical practices that seek to transform social conditions by any degree possible. While embarking on their moral missions, we find our participants \textbf{1)} navigate through deepening levels of embodied introspection to define their professional convictions, \textbf{2)} pragmatically weigh the tradeoffs of working within or circumventing institutional expectations to guide their technical approach, and \textbf{3)} look outside of themselves to nurture their practice. Our participants also demonstrate moral mission decision-making procedures when considering the use or refusal of Generative AI (GenAI) within existing responsible techniques. By engaging with~\citet{harney2013undercommons}'s theory of the \textit{undercommons}, we recognize the expansive potential of approaching RDS practice as a moral mission and informing design directions for RDS tools. Overall, our work contributes to AI ethics scholarship by centering the internal world of professional data scientists to surface moral decision-making procedures RDS tools should be built to support.
\section{Related Work}
\subsection{Values in Design and Responsible Data Science}
Although scholars have long recognized that technology can shape and be shaped by human values \cite{winner1980politics}, theories on how designers of technologies can and should actualize values in their work are ever-evolving. One of the first influential accounts of values in technological design is Value Sensitive Design (VSD), which defines a value as ``what is important to people in their lives, with a focus on ethics and morality''~\cite{friedman2006vsd}.
VSD employs iterative methodologies by integrating conceptual, empirical, and technical investigations to help designers navigate the relationship between values and technology.

However, the scope and nature of moral values remain an ongoing topic of discussion within VSD work: while a universalistic approach to human values risks imposing colonialist constructs onto local contexts, a relativistic approach risks condoning any practice as a valid expression of culture. Whereas earlier work from \citet{friedman2002human} endorses a conception of values as universal concepts that can have different instantiations in particular cultures, later work from \citet{borning2012nextsteps} rejects the universalist conception of values in favor of a pluralistic one. They also recognize that non-exhaustive heuristic lists of values (like those put forward by \citet{friedman2006vsd}) may nonetheless be valuable in practice.

The study of values in design is particularly relevant to RDS, a domain in which scholars have long highlighted the importance of situated and open-ended value inquiry \cite{morley2021ethics}. \citet{dignazio2020datafeminism}, for example, draw out seven principles of Data Feminism from feminist use cases of data science, including challenging power, embracing pluralism, and considering context. Other situated frameworks draw from feminist theories of justice \cite{ajmanietal2024dataagencytheory}, decolonial theories \cite{mohamedetal2020decolonialai}, and African philosophies \cite{barrett2025african} to highlight how legacies of power and collective experiences shape the curation and interpretations of RDS principles.

In addition to these frameworks, other scholars have also taken more universalist approaches to survey RDS values. Representative examples include AI4thePeople, a framework with five values created by \citet{floridi2018ai}. \citet{shen2025valuecompass} employ a set of 56 human values drawn from \citet{schwartz1994universalvalues} to measure value alignment between humans and LLMs across different cultures and scenarios. We see the rich theoretical scholarship of value-laden design and technical practice as powerful probes for reflection. Hence, we draw values from both relativist and universalist value frameworks mentioned above when designing value cards used in our interviews. 

VSD has also been critiqued for following what \citet{jafarinaimi2015values} call an ``identify/apply'' logic, which assumes moral problem-solving is conceptualized as a process of faithfully applying values that are already known to stakeholders. However, \citet{jafarinaimi2015values} object that this ``commonplace understanding of values in VSD as \textit{what a person or group considers important in life} [...] emphasizes the identification of values rather than their service in design situations.'' 
Although VSD acknowledges that conceptual investigation should be integrated with empirical and technical investigation, treating \textit{values as hypotheses }(coined by \citet{jafarinaimi2015values}) means that values are most comprehensively investigated within introspective, storied inquiries of practitioners' design experiences. Therefore, we center the storied inquiries of responsible data scientists to better understand the relationship between RDS values and practice.

\subsection{Responsible Data Science Techniques}
Researchers have developed numerous RDS techniques to address ethical concerns in AI development. Documentation tools have standardized how practitioners document metadata, data collection, methods, considerations, and intended use. Examples include Datasheets for Datasets \cite{gebru2021datasheets} and Model Cards \cite{mitchell2019model}.  
Fairness libraries such as \citet{bellamy2018ai}'s AI FAIRNESS 360 (AIF360) and \citet{bird2020fairlearn}'s Fairlearn are open-source toolkits to help practitioners detect, measure, and mitigate bias in their models. On the other hand, fairness libraries embed specific mathematical definitions of fairness that might not align with domain understanding of justice \cite{hertweck2024whats}. Explainability techniques aim to make model behavior more understandable by targeting transparency~\cite{lipton_mythos_2017}, though critics argue interpretability does not guarantee stakeholders can act on explanations or that transparency alone shifts power dynamics~\cite{selbst2019fairness}.
Practitioners also regularly use less formalized data cleaning, demographic balancing, and anonymization techniques~\cite{hsu2021open}.
All of these techniques can degrade into performances of ethics that fit within organizational logics and misdirect efforts for meaningful change \cite{metcalf2019owning}. 
 
Interactive tools developed in HCI research aim to make ethical deliberation more accessible and participatory. Cards and gamification have been used as approaches to navigating ethics in HCI research \cite{friedman2006vsd,widder2024power}. An example is Value Cards \cite{shen2021value}, which helps teams navigate through conflicting values during the design process. Bitácora is built for nonprofits using social media data; it recognizes that organizational context matters for what counts as responsible~\cite{alvarado_garcia2024bitacora}. 
Checklists also aim to provide practitioners with clear and actionable guidance \cite{madaio2020co-designing}. 
While these approaches give structured time to explore ethical dilemmas, they all require collective time \cite{lin2022learning,murray-rust2023grasping}. 

A further challenge for RDS tools pointed out by \citet{wong2023seeing} is that most tools are siloed, addressing isolated problems rather than integrating into actual workflows, leaving practitioners to figure out which tools matter and whether they can use them at work. Crucially, these tools tell practitioners what to evaluate but offer little guidance on how to understand their own positionality when taking value-laden sociotechnical actions. The rise of GenAI also exposes limitations in existing RDS techniques. \citet{wolfe2024implications} show that GenAI introduces new problems, including attribution, consent, and environmental impact, that existing tools were not designed to address.

\subsection{Evaluating Responsible Data Science Practices}
Prior work in operationalizing RDS has largely focused on proposing and evaluating lifecycle-oriented frameworks, capturing the challenges of toolkit adoption, and surfacing workflows and perceptions from practitioners \cite{zhang2020how,heger2025towards,krijger2023ai}. RDS frameworks outline the moral concerns that arise across the entire data science process, from problem formulation and data collection to modeling, evaluation, deployment, and communication \cite{saltz2019data,stoudt2024ten,urovi2024taps,boenig-liptsin2022data}. To understand the technical needs and feasibility of RDS, prior HCI scholarship has explored the efficacy of existing RDS toolkits \cite{deng2022exploring}. Works from \citet{varanasi2023currently, bhattacharjee2025technical} acknowledge how institutional structures impact the ability for AI and ML practitioners to uphold responsible AI values. The challenge grows as corporations increase adoption of GenAI in engineering workflows. 

Emerging work further aims to surface the introspective beliefs, deliberations, and strategies of RDS to inform a more holistic and generative understanding of RDS practice to better support it \cite{bennett2025artificial, madaio2024tinker}. \citet{dong2026evaluating} offers a useful lens for addressing this gap by reframing RDS as a problem of practitioner behavior change. Rather than evaluating responsibility only through models, datasets, metrics, or organizational policies, responsible outcomes also depend on the everyday behaviors and decisions of data scientists \cite{dong2025behavior}. We build on these prior works by emphasizing that evaluating RDS ultimately requires understanding how practitioners repeatedly enact those values in their concrete decisions.
\section{Methods}
Our study was approved by our institutional review board and pre-registered with the Open Science Foundation.\footnote{\url{https://osf.io/g56fd}}
\begin{figure}
    \centering
    \includegraphics[width=1.03\linewidth]{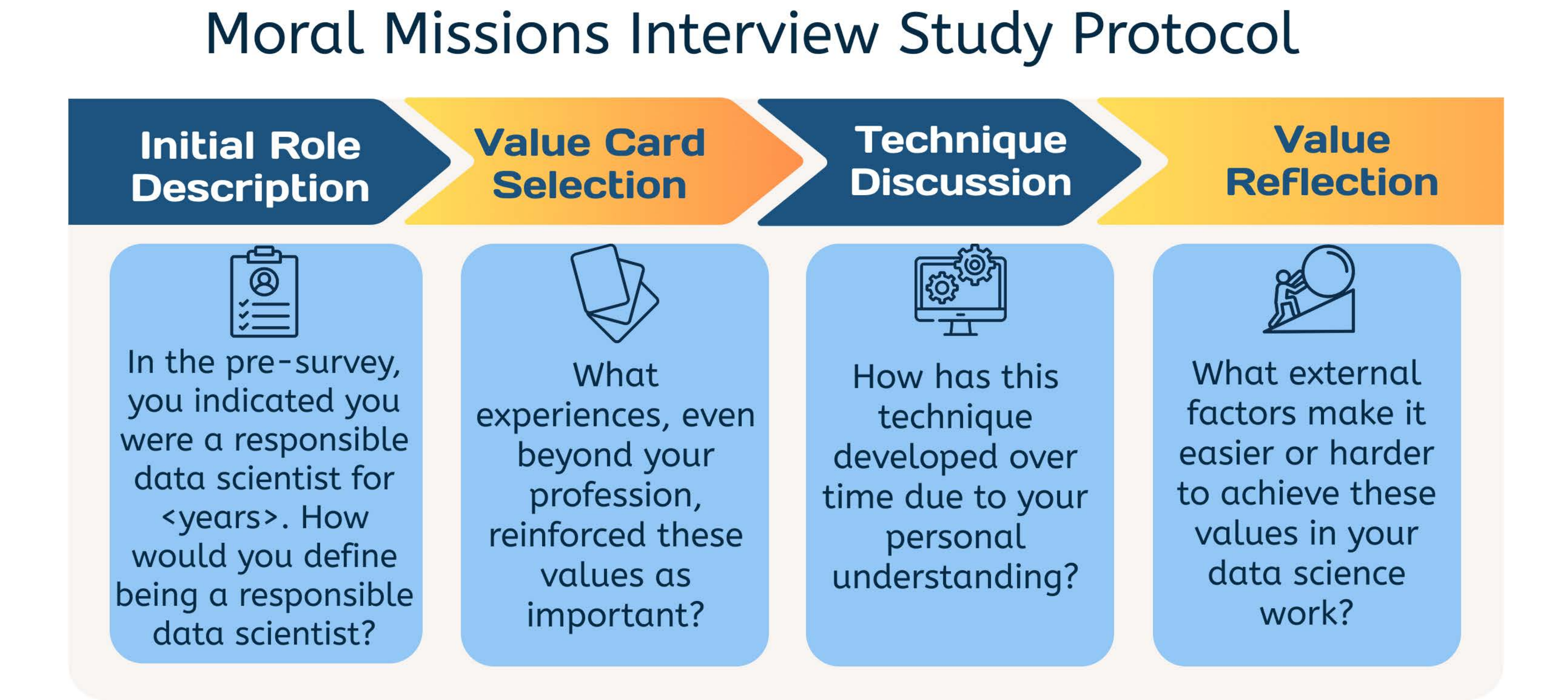}
    \caption{Semi-structured interview used to elicit participants' RDS values, selected techniques, and reflections on the conditions that shape value enactment in practice. Example questions from the interview protocol are featured at each stage.}
    \label{fig:study}
\end{figure}

\subsection{Participants}
For this study, we aimed to recruit participants who identify as RDS scientists or AI practitioners in industry, government, and non-profit sectors. We recruited participants through RDS Slack channels, LinkedIn posts, local RDS meet-ups, and word of mouth throughout the research team's network. All participants completed a pre-survey (refer to page \pageref{sec:pre-survey} for details) to detail their data science experience and focus, identify an RDS technique to discuss in the interview, provide optional demographic information, and finally confirm their consent and accessibility accommodations (breaks, captioning, access to questions beforehand). In the end, we recruited 15 participants from diverse industries, experience levels, and data science roles. The interviews on average lasted 82 minutes, and participants were compensated \$25 for their time. An overview of our participant demographics is presented in Table \ref{tab:participant-characteristics}.

\begin{table*}[!ht]
\centering
\resizebox{\textwidth}{!}{%
    \begin{tabular}{lllllll} \toprule
    \textbf{Alias \textit{(pronouns)}} & \textbf{Role} & \textbf{Industry} & \textbf{Data Science} & \textbf{Country of} & \textbf{Ethnicity} & \textbf{RDS Technique} \\
    & & & \textbf{Experience (Yrs)} & \textbf{Residence} & & \\ \midrule
    P1 (\textit{they/them}) & Battery Scientist & Corporate & 8 & X & Asian & Up \& Down Sampling \\
    P2 (\textit{she/her}) & Data Scientist & Government & 5 & United States & European or White & Pre-mortem Documentation \\
    P3 (\textit{X}) & Data Scientist & Government & 10 & X & X & Removing PII \\
    P4 (\textit{she/her}) & Product Manager & Corporate & 12 & United States & European or White & Model Documentation \\
    P5 (\textit{he/him}) & AI Practitioner & Non-Profit & 3 & Nigeria & Black & Human-in-the-loop Advocacy \\ 
    P6 (\textit{she/her}) & Machine Learning Engineer & Corporate & 3 & Canada & Black & Model Cards~\cite{mitchell2019model} \\
    P7 (\textit{he/him}) & Data Scientist & Government & 3 & United States & Asian & Datasheets for Datasets~\cite{gebru2021datasheets} \\
    P8 (\textit{X}) & Research Analyst & Non-Profit & 3 & United States & Black & Removing PII \\
    P9 (\textit{she/her}) & Developer & Non-Profit & 5 & Canada & European or White & Serverless Data Governance \\
    P10 (\textit{she/her}) & Modeler & Non-Profit & 5 & Nigeria & Black & Model Calibration Documentation \\
    P11 (\textit{she/her}) & Data Scientist & Corporate & 6 & X & Asian & SHAP~\cite{lundberg2017unified} \\
    P12 (\textit{he/him}) & AI/ML Engineer & Non-Profit & 4 & Nigeria & Black & Subject Matter Expert Consultation \\
    P13 (\textit{he/him}) & Principal Data Scientist & Corporate & 18 & United States & European or White & Distributive Visualizations \\
    P14 (\textit{he/him}) & AI/ML Engineer & Corporate & 5 & Nigeria & Black & Reducing API Use \\
    P15 (\textit{he/him}) & Marketing Data Engineer & Corporate & 6 & Mexico & X & Encrypting PII \\
    \bottomrule \end{tabular}
}
\caption{A table summarizing characteristics of each 15 participants, including pronouns, data science role, industry, years of data science experience, ethnicity, country of residence, and the responsible data science technique they discussed in their interview. Note: optional information participants chose not to disclose indicated with an X.}
\label{tab:participant-characteristics}
\end{table*}

\subsection{Interview Study Design}
To gain insight into the lived experiences of data scientists using RDS techniques, we designed a semi-structured interview with an interactive visual probe. The visual probe was a workspace in which participants could annotate, interact with pre-made value and technique cards, and add online media. The interview protocol was split into four phases: initial role description, value card selection, technique discussion, and finally, a value-grounded reflection of their practice (refer to Figure \ref{fig:study} and for more details refer to page \pageref{sec:interview}).

\textbf{Data Science Role Description.} In the initial role description phase, participants described their current role as a data scientist. We asked participants to describe the data they work with, the products, systems, or initiatives they contribute to, and the colleagues, communities, or stakeholders they interact with. 

\textbf{Data Science Value Card Selection.} After providing an overview of their work, we introduced the participant to the interactive visual probe workspace. Participants were given time to familiarize themselves with the interface and then asked to select three to five value cards they personally deemed important to their work as a responsible data scientist. The value card form and selection activity are inspired by VSD Envisioning Cards~\cite{yoo2022multi} and AI ethics education value cards~\cite{shen2021value}. The provided value cards were curated from universal design value research and ethics principles informed by specific epistemic communities: Data Feminism~\cite{klein2024data}, AI4thePeople~\cite{floridi2018ai}, African Data Ethics~\cite{barrett2025african}, and Disability Justice principles~\cite{berne2018ten}.

\textbf{Responsible Data Science Technique Discussion.}
After selecting of value cards, we moved to the RDS technique discussion. We defined RDS techniques as technical practices within the data science workflow to improve, consider, mitigate, or evaluate social implications of data-driven technology. Participants were asked to identify a technique to discuss for the duration of the interview. We also provided technique cards to inspire participants based on reviews of RDS tools ~\cite{dong2025behavior,kuehnert2025who}. In the discussion of the technique, we asked them to identify value cards relevant to their technique of choice, how they learned about the technique, use cases, and process to implement the technique. 

\textbf{Value Reflection.}
Finally, we asked the participants to reflect on their technique through the lens of their values. Participants discussed how their values were actualized in practice, what external factors influenced this actualization, and how their larger professional community should engage with RDS.

\subsection{Data Analysis}
The data were primarily collected through transcribed and recorded interviews. Additional data were collected through the pre-survey, interview notes, and the visual probe artifact. All of the participants’ anonymized and cleaned data were categorized in folders with de-identified participant ID labels. 
Participants with identifiability concerns were offered the opportunity to review relevant data (i.e., recordings and transcripts) and select items for anonymization or removal. Participants were also informed that they could exit the study at any point, even upon study completion, and could ask for relevant collected data to be excluded.

\paragraph{Qualitative Analysis Methodology.}
We developed an analysis plan with a foundation in interpretative phenomenological analysis (IPA) ~\cite{dahlberg2020open}. Phenomenological qualitative analysis aims to explore lived experiences by centering the individual’s ability to articulate a comprehensive account of a meaningful activity ~\cite{eatough2017interpretative}. Phenomenological analysis has been deployed in HCI studies to gain a deep understanding of user needs, existing technology use, and design implications ~\cite{linder2017ipa, kallia2025to, midha2022lived, horvath2025how, fortin2013beyond, rizvi2025i}. Its strengths are especially apparent in capturing novel, introspective activities such as moral decision-making in RDS practice. Each participant’s understanding of their personal and social world grounds analysis and interpretation in individual nuance often smoothed over in thematic analysis~\cite{midha2022lived}.

\paragraph{Analysis Approach.}
A subset of the research team (two interview facilitators and a non-interviewing researcher) performed the qualitative analysis using Atlas.ti. Data were evenly distributed among the team. We conducted three rounds of analysis to construct a phenomenological account of each participant’s process of integrating and adapting their example of an RDS science technique. First, the team reviewed each participant’s data sources multiple times to facilitate familiarization. During this phase, one‑page documents were created for each participant to capture potential design considerations arising from their experiences. The team then coded transcripts and visual probe artifacts to identify aspects of interest. We developed narrative accounts for each interview to document our interpretation of how participants’ experiences addressed our research questions, and we met as a team to review them collectively, apply theoretical concepts, and reach interpretive consensus. Finally, we drew on the resulting codes, narratives, design considerations, and theoretical touchpoints to generate themes, an overarching conceptual model, and design implications \cite{naeem2023step}.
\section{Findings}
\begin{figure}[ht]
    \centering
    \includegraphics[width=1\linewidth]{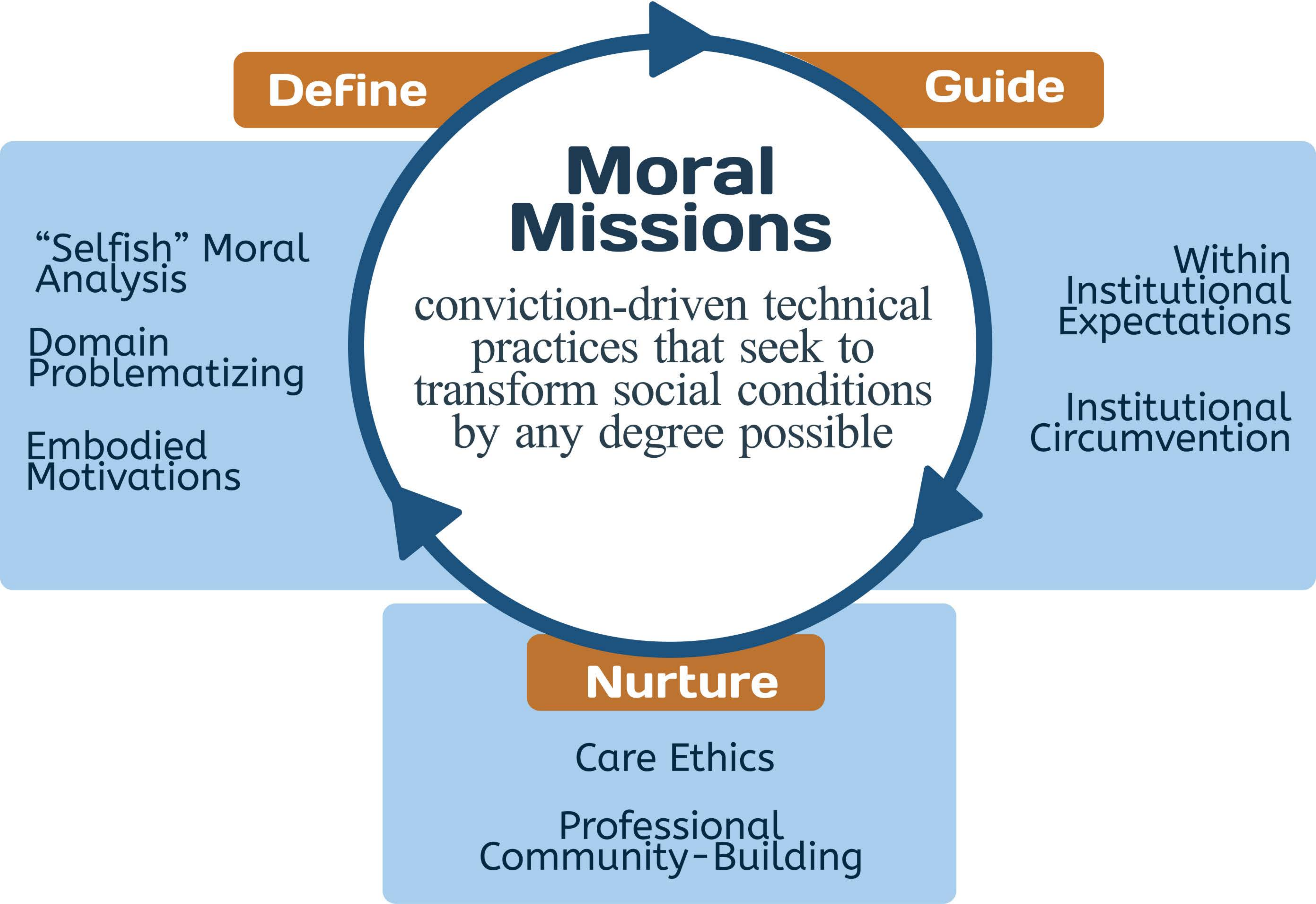}
    \caption{Conceptual model of moral missions as an iterative process of defining moral commitments, guiding RDS practice under institutional constraints, and nurturing commitments through relational and professional communities.}
    \label{fig:moralmissions}
\end{figure}

We analyze our fifteen participants' moral decision-making procedures through discussions of personal data science values \textbf{(RQ1)} and accounts of implementing RDS techniques \textbf{(RQ2)}. Our analysis demonstrates the phenomena we call \textit{moral missions} are: 1) defined through embodied introspection, 2) guides for innovative circumventions of institutional rigidity, and 3) nurtured by relationality beyond the workplace. To close, we highlight a case study of these concepts in action with the emergence of GenAI.

\vspace{\baselineskip}\noindent\textbf{RQ1: How do data scientists committed to building responsible data-driven technology identify their personal values?}

\subsection{Defining Moral Missions: Embracing Embodied Introspection for Conviction Development}
At first, many participants were uncertain or hesitant about articulating their moral priorities. Once they engaged with our interview probes, participants revealed layers of moral reasoning that ultimately led to intimate and emotionally grounded motivations for their practice. 

\subsubsection{``Selfish'' Moral Analysis}
We asked participants to select a subset of value cards they deem most important to their responsible practice. Several participants explained they often don't have the space within their workplace to explicitly discuss their values (P1, P2, P6, P10, P15). Therefore, many participants exhibited reservations about exercising moral agency. For example, P2 qualified their creation of the value \textit{Collective Well-Being} with selfishness: 
\begin{quote}
    \textit{I think I tend to be pretty empathetic as a person. So it's almost selfish; the sense that other people are doing okay around me [...] is very comforting. I mean, that's sort of the selfish angle; I'm comforted by other people.}
\end{quote}

A subset of participants who lead responsible data-driven organizations were more assured of their moral agency and readily bypassed our proposed values to present their own (P9 and P13). For instance, P13 created a value card for \textit{Epistemic Responsibility}:
\begin{quote}
    \textit{Epistemic responsibility is one view of an accurate, ethical way to show data. [...] I think our greater responsibility is not saying accurate things; it's promoting accurate beliefs. This is maybe related to consequentialism, but thinking about the effects, how things are perceived, and taking responsibility for a few more steps down the chain of communication.}
\end{quote}

\subsubsection{Domain Problematizing}
Across all participants, values were jumping-off points to problematize and subsequently construct moral analyses of their data science domain. For example, P1 models electric vehicle (EV) battery performance based on battery data across EV manufacturers. In selecting \textit{Consider Power}, P1 articulated how American companies and their powerful supporters determine data access and, in turn, shape domestic EV purchasing decisions in their favor.
\begin{quote}
    \textit{Teslas are the majority of used electric vehicles out there, and a lot of that has to do with the amount of influence and power that company has been in successfully selling their products. There are a lot of very well-made Chinese models that do not exist in the US because of import rules. There are a lot of [original equipment manufacturers] (OEMs), manufacturers, core sellers in the US that do not want Chinese brands into the US. So that influences the amount of data that we have and what vehicles we can serve.}
\end{quote}

While considering \textit{Freedom from Bias}, P15 realized his work in targeted marketing contradicted his moral inclination.
\begin{quote}
    \textit{On the marketing side, you just want to create the audiences, so targeting, exclusiveness, or going in that direction would be what people are looking for. [...] I mean, at the point of the arrow, the data references should be [...] clean and free of bias [...] I mean, I will not [select `Freedom from Bias'] but I just want to bring this idea out of my head.}
\end{quote}
Later in the interview, P15 was asked to consider what it would look like to practice targeted marketing free from bias. He acknowledged he had not considered this tension before, but imagined regulations on advertising frequency were in the right direction. Across these instances, problematizing through values surfaced implications of participants' work and compelled them to take a position.

\subsubsection{Embodied Motivations to Pursue Data Science}
As the discussion of values continued, participants moved beyond abstract morality or domain knowledge to share the intimate experiences that established their emotional connection to RDS~\cite{klein2024data}. For example, P8 and P12 returned to their hometowns. P8 grew up without many resources and did not appreciate the "savior complex" non-African NGOs adopted to fundraise for policy interventions. He assumed his role as a policy data analyst to inform narratives that respect the dignity of underresourced communities. P12 was born in North-Central Nigeria and grew up watching people around him struggle. When he found out about multilingual language models, he saw it as a powerful means to improve his community's conditions.
\begin{quote}
 \textit{I seek to contribute to the lives of people, probably starting with just one person. Someone who is smart but struggles, whose parents cannot cater for them. I'm committed to helping improve the quality of life of people [...] The whole idea is how can locals in the interior parts of the continent access the same technology that people in the diaspora and in the urban areas enjoy.}
\end{quote}

P2, P4, and P11 were drawn to RDS by their personal experiences as women. P11, particularly as a woman of color, observed how systems---AI included---can treat people unfairly based on identity:
\begin{quote}
    \textit{Having grown up as a woman of color, [I have seen] when these biases and stereotypes people have are enforced in different situations and how harmful that can be, and how much upfront unfairness a lot of people suffer. So this has very much informed my work. With my personal lived experiences, AI and machine learning [should] not include these factors at all, instead of saying: oh no, we are using gender, but we are not really using it to discriminate. How do you know? How much control do you have over an AI or ML model that you can for sure say that?} 
\end{quote}

Most participants cited formal and informal training as formative to internalizing the importance of their responsible practice. P14 shared how a single conversation with a colleague shifted his perspective on third-party API use:
\begin{quote}
    \textit{Two years ago, I was working on a project with a colleague who is also an AI engineer. One of the discussions we were having about a project, he spoke about the fact that [...] users would have to enter information about their therapies and things like that, which is quite private. He highlighted the fact that we would have to reduce third-party API use as much as possible, because these are private information and the third-party developers of these APIs can't have access to this information [...] I'd always known that third-party APIs would more likely have access to the information we bring in, but I've mostly not cared. If I get my work done and it does the job, just go ahead and use it. But then having that conversation opened my eyes to seeing that we have to be more careful.}
\end{quote}

\vspace{\baselineskip}\noindent\textbf{RQ2: How do data scientists committed to building responsible data-driven technology adapt and integrate responsible data science techniques into their work?}

\subsection{Moral Missions as Guides: Navigating Institutional Expectations} 
Participants' moral missions guided them to make the most responsible decisions within their institution's expectations or circumvent institutional logics to maintain their convictions.

\subsubsection{Working within Institutional Expectations}
Participants ran into challenges at several levels. At the institutional level, responsible practices often lose out to business priorities. For example, P4 described how model and data documentation were never prioritized by the leadership:
\begin{quote}
    \textit{From a business perspective, it's not ever going to be the highest priority item to solve. The bottom line is profit is the number one driver until someone breaks something. Then it's the highest priority to solve, but it's usually not thinking long-term.}
\end{quote}

Working within these institutions often involved acknowledging the current limitations of the data or systems the practitioner was working with. Participants like P1, P4, P10, and P11 did their best to communicate sociotechnical limitations to their non-technical colleagues to exercise a form of moral agency in their work. P1 described flagging limitations to product and marketing colleagues before decisions were made: 
\begin{quote}
    \textit{I try to keep in mind, like, oh, these are the biases that we're dealing with. And these are the limitations, so maybe don't trust this result as much. Even though the stat says there's a 95\% probability that these two groups are different by X amounts. That's without the context of how I trust the data. So then, with my input, it's like, oh, maybe you don't make a decision based on this.}
\end{quote}

When evaluating the efficacy of working within their explicit institutional role, participants who felt limited by their institution speculatively turned to policy change and regulations to expand their opportunities for responsible practice (P1, P5, P15). These participants' hope in policy was confirmed by P11, who was not worried about RDS losing priority within her institution, even with the rise of GenAI:
\begin{quote}
   \textit{When you have regulatory oversight, they cannot [deprioritize explainability], because it's not dependent on them. Regulators are probably the biggest engine of change I have seen with respect to moving towards explainability.} 
\end{quote}

Participants also described negotiating with funders, clients, policymakers, and government officials to reconcile differing priorities and definitions of responsible practice. P10 models malaria public health interventions to predict which local government areas would benefit the most. She describes how she presents her model to program managers as a support tool for equitable decision-making:
\begin{quote}
    \textit{We always let the program manager [know that] modeling is a decision support tool. These are things which help them to at least have some evidence to objectively make decisions. Maybe [previously, they] based their decisions on [their] state of origin. [...] If you channel the resources you have [for] maximum impact, incrementally the margin of reduction [of malaria transmission] will be more.}
\end{quote}

\subsubsection{Strategies to Circumvent Institutional Expectations} 
For some participants, institutional pushback expanded their moral missions to devise innovative sociotechnical strategies (P2, P12, P13).
P9 held her team's position even when it cost her a client: 
 \begin{quote}
   \textit{There was a question from the government about where our servers were located to protect the data privacy [...] it took us a couple of weeks to realize we didn't have that much data, and we didn't need to store data. We really didn't need to. So then we'd gone back to the government and said, 'we don't have any servers; we're not getting any.' And they wouldn't work with us anymore, because the government had a rule that said that the data had to be safeguarded in your own server, in your own data container [...] We were sticking to our rules, our values about what we were going to do.}  
 \end{quote}
P13 completed a project ``out of spite'' to prove his novel, equity-focused technique can rival traditional visualization approaches.
\begin{quote}
    \textit{I had another client that was actually social advocacy focused, and they were like, `no, no way, we're gonna do the traditional thing, that's what's gonna sell it to funders.' That was weird and frustrating, but it also kind of triggered this; I'm pretty sure I can prove you wrong. I want to prove that this is real. From the startup world, I had a sense of how to run an A-B test and all that. So, [I] just started testing some of the ideas to see [whether] certain ways of presenting disparity data lead to stereotypes or lead to victim blaming. It was a thing that was speculated online, but there wasn't anything empirical on it. So, mainly out of spite, I ran those experiments to just see what would happen. It turns out that, yeah, it does have an effect and got results on that.}
\end{quote}

In more hostile institutional environments, such as P2's, surface-level compliance created a cover for technical resistance: 
\begin{quote}
    \textit{One of the ways that I've seen models around by other people in the civil service has been continuing to do the work, but just calling it something else [...] There's a lot of integrity that I saw of people just, like, well, this is the type of data that we know people find useful, so we're gonna keep doing the work. As the [executive orders] (EOs) were rolling out, and they're like, can't talk about women--- okay, fine, just take the word women down [from the documentation].}
\end{quote}

\subsection{Nurturing Moral Missions: Relationality and Collective Decision-Making}
Regardless of the decisions made by participants, they knew they needed to consider relationality in their technical actions to take their moral missions to the next level~\cite{birhane2021algorithmic}. Throughout their accounts, participants made a point to acknowledge end users, impacted communities, or supportive professional communities.

\subsubsection{Care Ethics}
For many of our participants, the most crucial decision-making parameter was the impact on the end user or data subject population. Several participants' practices resonated with theories of care ethics, which prioritize personal and concrete relationships with others \cite{zegura2018care}. These participants made decisions by thinking about end users and the larger communities their work could harm or help. 
P9 described the care her team put into data handling as coming from a deeply personal sense of accountability: \textit{``We would have felt so bad if that data had been intercepted. We did not want to be the ones who caused that type of breach.''}

P12 held a similar line when it came to speed: \textit{``Speed is a factor. But it cannot be done at the expense of good data. We can't choose speed over credibility.''} P12 grew from working with a single native speaker to collaborating with a team of on-the-ground data collectors because he understood the scale required to responsibly represent low-resource languages: 
\begin{quote}
    \textit{I began to work with just one person, but again, I understood that this could be done at scale [...] if you want to build really good technology, you need to represent diversity. And if you want to build anything at scale, then you need to be able to collect a whole lot of data at approximately the same time.}
\end{quote}

\subsubsection{Professional Community-Building}
Participants working within their institutions took time to build collective understanding over time (P1, P3, P4, P7, P8). P1 felt the potential to push against the data bias they cared about when they talked with their colleagues outside of work:
\begin{quote}
    \textit{I have an inkling that my colleagues agree with me, but that's just because I know what their other values are based on other conversations [...] But in the professional environment, it's more of like, oh, we just talk about these as facts.}
\end{quote}

P15's data science and engineering teams developed a seamless, privacy-preserving, and responsible technical solution with roundtable discussions.
\begin{quote}
    \textit{[Given the] concerns on what we should do with the IPs, somebody said, ‘okay, if we use this process on hashing the IP, we'll get different values for each of the entries. We'll be able to count each of them independently, and will not have this compliance situation exposing the IPs.' He was a data engineer. He encrypted the IP, and once the table was ready, I performed some analysis on counting the different entries for the different timestamps and countries. I ran the model expecting to have the same marketing attribution rate.}
\end{quote}

As P2's workplace became increasingly hostile while being furloughed for six weeks and losing access to work communications, she found a crucial lifeline to communicate with colleagues outside of work.
\begin{quote}
    \textit{I think everyone felt anxiety, and some grief, and maybe some depression; I don't know. One thing that was great, was just being able to keep in touch with coworkers over things like Signal so that we have a space to communicate and support each other, and we don't have to worry about surveillance or something.}
\end{quote}

\section{GenAI: An Emergent Moral Missions Case Study}
The rapid corporatization of GenAI tools has introduced distinct market pressures that fundamentally complicate the moral missions our participants described~\cite{lorrimar2026silicon, challenger2025jobs,ombAI2025}. More than half of our participants raised GenAI as an emerging circumstance that evolved their moral missions. Unlike academic contexts, where communities of established and well-respected professionals openly dissent from unrestrained AI adoption~\cite{bender2025ai}, software engineers and technical practitioners in industry operate within incentive structures that actively reward, and in some cases, compel AI use~\cite{wolfe2024implications, connolly2024public,tinsman2026navigating}. 

\subsection{Defining GenAI Moral Missions}
Even if participants didn't use GenAI for their responsible practice, most noted the unavoidable market push of GenAI across sectors. As P6 put it:
\begin{quote}
    \textit{Whenever I get a new job, which has mostly been internships, they usually ask you to look at the organizational structure and think of what you can work on. [...] A lot of companies are not very sure of what they want to actually achieve with ML, and now we have LLMs. Many people are actually not sure. Well, a few people are sure. When they pull up RAG, at least you know they want to search a particular document or something. Sometimes there's a grey area, and they want experimentation.}
\end{quote}

While some participants embraced GenAI (P7, P11, P14), others explicitly questioned the hype. P4, a product manager collaborating with data scientists, shared her thoughts on the ongoing enthusiasm in her workplace to adopt GenAI. 
\begin{quote}
    \textit{My work opinion is [to be] cautious but experiment. My personal opinion is similar in that [...] be very, very cautious and experiment with the understanding that the data, the developmental process, to me, for these [GenAI technologies] were not ethical. No one had informed consent as to whether their stuff they put out on the internet was going into these; no one was paid for it, and no one even had the option to opt in or out.}
\end{quote}

\subsection{Moral Missions Guiding GenAI Use}
To P7's surprise, institutional interest in GenAI's potential for increased efficiency bolstered support for his effort to improve dataset documentation. Due to P7's measured GenAI approach, he not only exceeded institutional expectations but also met his responsibility standards.
\begin{quote}
    \textit{When the new administration was coming in, we thought that it would change everything, but it was quite the opposite. [...] They actually featured our dataset on their website and said that we are trying to make this process much faster and better with quality checks. It's not like we are making [it] faster, but we are dropping some things, or we are skipping some things; that's not the idea.}
\end{quote}

P13 did not personally use GenAI for his technique, but technical gaps in popular data visualization tools like Power BI led to his reluctant support of code generation. He instructs novice coders to: \textit{``go to Gemini, ask it to create a D3 page showing this data. Don't trust anything that it spits out, but check it, make sure that it's right.''}

P4 described staying engaged with workplace discussions of GenAI adoption, despite her own reservations, as a deliberate choice to stay in the decision-making room: 
\begin{quote}
    \textit{I still think that I have an obligation to test and do things and try to use them. Because the people that are staunchly against it and very vocal about it are not the ones invited into the room where decisions are being made for a business on whether or not to use them. But I'm also not going to be gung-ho, put it in everything, do everything like a lot of people.}
\end{quote}

With expertise in explainability, P11 explains how GenAI is technically misaligned with their approach to RDS practice.
\begin{quote}
    \textit{Previously you would have a small set of data, then you would use pre-selected features, and then you would use it to predict something. So you had a lot of control over it. Whereas when we talk about LLMs based on transformer architecture, companies like OpenAI and Anthropic, they're literally using the whole internet. You just don't know where it's learning from [...] with LLMs, explainability, while critical, has become a really big challenge, because you just don't have a single source of information or corpus from which the information has been pegged.}
\end{quote}

\subsection{Center Relationality to Nurture GenAI Moral Missions}
Regardless of the decision to adopt or struggle against GenAI use, participants recognized their responsibility to maintain human oversight and engage their communities to scale responsible GenAI efforts. P7 built a pipeline that kept human judgment throughout the process: 
\begin{quote}
    \textit{We have developed a pipeline where we can ingest those PDFs. I first extract the text from that, and then use different LLMs, like Gemini 2.5 Flash, to externalize: who is the main federal agency responsible for this work? What is the location? We have developed some prompts to extract those and validate. [...] The initial annotation is human, then improving and scaling it up to maybe [millions of] pages is automatic. And then the final evaluation [is] again human. So, it's kind of like two places where humans are there.} 
\end{quote}

P5's entire moral mission is equipping his local RDS community with the knowledge to advocate for African representation in GenAI. Therefore, he regularly engages with the technology as a method for identifying African exclusion and was invigorated by the growing movement he plays a part in.
\begin{quote}
    \textit{I'm talking about [GenAI in] my region; people are becoming more interested in [AI ethics]. It's not like before; people didn't really put much interest there. The awareness is taking momentum, and people are really coming into it and knowing that, okay, I have to speak, I have to advocate, I have to be included. So, it's making the whole thing easier now. I expect good inclusiveness in the future.}
\end{quote}

Across interviews, participants described GenAI not simply as a new technical tool, but as a site where organizational urgency, market pressure, professional identity, and moral conviction collided. Participants in our study did not all respond to GenAI the same way, with some refusing to use it altogether. Framing refusal as part of deeply held moral missions reveals that such practices are not simply \textit{subtractive} stances against GenAI, but rather \textit{generative} practices opening up new forms of relationality outside academic and capitalistic norms \cite{tuckandyang2014refusal}.
\section{Moral Missions, the Augmented Undercommons, and Implications for Responsible Data Science Communities}
The seemingly irreconcilable relationship between RDS practice and institutional logics is a persisting critique we surfaced in our study as well \cite{metcalf2019owning, wong2023seeing, deng2022exploring}. We are the first to turn to the Black Studies theory of the \textit{undercommons}~\cite{harney2013undercommons} to flag clearer directions for decoupling RDS practice from institutional logics. \citet{harney2013undercommons} articulate how professionals within academia subvert the often-exploitative conditions of their workplace to exercise their moral agency. While~\citet{harney2013undercommons} first theorized the undercommons for the academy,~\citet{croskey2026augmented} introduced the augmented undercommons as the first theoretical work to bring these ideas into HCI, describing how subversive designers operate within and against institutional constraints to advance justice-oriented work. We extend this theoretical lineage to illustrate how moral missions call for subversive data scientists, who \textbf{refuse credit, operate in parallel, and center collectivity} as the foundation of responsible practice \cite{croskey2026augmented}.

\subsection{Refusing Credit: Define Moral Missions}
The theme of \textit{Domain Problematizing} reflects the idea of refusing credit as explored by \citet{harney2013undercommons} in the undercommons. The concept of credit represents the feeling that professionals `owe' full cooperation to the institutions that employ them. In `paying off' this credit, professionals tend to act in the best interests of their institutions (i.e., RDS appeals to reducing business risk \cite{wong2023seeing, metcalf2019owning}) even as these institutions uphold systems that harm and push communities to the margins. Participants struggled to exercise their moral agency because they had institutional expectations in the back of their mind.

\citet{harney2013undercommons} contend that \textit{refusing credit}, or letting go of the feeling of owing cooperation, can foster ``fugitive publics'' that ``\textit{operate outside the logic of institutional credit and repayment.}'' \citet{madaio2020co-designing} similarly identify ``good tension'' as a necessary catalyst for meaningful responsible AI discussions. When data scientists are able to view their professional responsibility as not in service to institutional ends but to societal progress, it may become easier to not only develop personal convictions about one's professional domain but also to act on these convictions.  

\subsection{Operating in Parallel: Guide Moral Missions}
Although working within institutional priorities can build buy-in for responsible techniques~\cite{madaio2020co-designing}, participants in our study recognize that this approval of their responsible practice comes at the cost of fully realizing their moral missions. They therefore \textit{operated in parallel} by simultaneously translating community needs into professionalized spaces and challenging the boundaries of those same spaces~\cite{croskey2026augmented}. Materially, participants risked losing clients (P9) and even discussed professional dissent outside of work (P1 and P2), establishing clear and often risky stances on the importance of centering communities within AI development regardless of the economic and social cost of such a stance.

\subsection{Center Collectivity: Nurture Moral Missions}
Our theme of \textit{Professional Community-Building} also makes clear that refusal is hard to sustain alone. \cite{harney2013undercommons} emphasize that building power requires study and planning in community, with colleagues, fellow professionals, and collaborators who share the stakes. When participants felt like the only ones concerned about moral dilemmas in their work, they embodied the principle of \textit{collectivity} by forming connections with other subversive data scientists~\cite{croskey2026augmented}. Collectivity is not only about sharing critiques but developing ad hoc strategies ~\cite{lin2026heres, so2025cruel, abdalla2025robot}. In line with the idea of the undercommons, sometimes these solutions can only be shared in privileged spaces and between colleagues with established relationships, but other times, the open sharing of RDS techniques is the catalyst for new fugitive publics. 

\section{Design Implications for RDS Tools}
Our findings suggest that RDS tools should support not only the application of ethical principles, but also the moral decision-making procedures through which practitioners sustain those principles under institutional constraint. We therefore identify design implications for sociotechnical systems that help practitioners define, guide, and nurture moral commitments throughout data science work (see prototypes of design implications on page \pageref{sec:example}). 

\subsubsection{Develop moral agency}
Previous scholarship recommends embedding learning resources inside of RDS tools to support practitioners' understanding of data ethics frameworks and theory so they can be appropriately applied to practice~\cite{deng2022exploring, wong2023seeing}.
However, inspired by~\citet{jafarinaimi2015values} and the rich moral analyses developed by our participants, we contend that even simple data ethics prompting can be powerful jumping-off points for ethical contextualization.
The success of existing card-based tools supports this direction but often focuses on developing analyses on behalf of stakeholders \cite{friedman2012envisioning, shen2021value}. 

Instead, we imagine tools that build convictions and emotional investment for responsible practice may improve practitioners' capability to contextualize across projects~\cite{dong2026evaluating, madaio2020co-designing, madaio2024tinker, urovi2024taps}. 
These tools can look like adapting existing value cards to probe personal moral analyses, appropriating short structured templates like positionality statements into journaling prompts for career-long introspection, or assessing practitioners' alignment to existing data ethics frameworks for meaningful investigation.

\subsubsection{Maintain sociotechnical strategies}
Previous critiques of technical RDS toolkits highlight how generalizable techniques can be co-opted into ethics-washing and reinforce technosolutionist orientations that avoid the social dimensions of responsible practice~\cite{selbst2019fairness, madaio2020co-designing, wong2023seeing}.
To combat the misappropriation of RDS techniques, previous scholarship called for embedding contextualized examples within toolkits and building avenues for non-technical stakeholders and community members to introduce socially-grounded context to technical considerations~\cite{deng2022exploring}.

Our study reinforces critiques of purely technical RDS tools. Practitioners provided technical specifics in interviews, but the richest aspects of their technical narratives were the social considerations they made along the way; for example, determining stakeholder relationships to maintain privacy vision (P9), preparing communications to inform colleagues of responsible considerations (P1, P10, P11), and applying responsible scholarship to their practice (P7, P13). All of these demonstrate how technical and social action are deeply intertwined. 

In addition to curating examples~\cite{bako2023understanding}, we should develop tools to store their instances of responsible practice. Similar to repository-based platforms like GitHub, Kaggle, or Hugging Face, practitioners should be able to store sociotechnical artifacts of their responsible practices~\cite{zhang2020how}. They can also keep track of what went well or what went wrong in each instance to retain and refine successful strategies to carry on to new contexts~\cite{urovi2024taps}.
Explicit access control should be a major feature of these repositories, especially if these strategies are subversive. Practitioners should be able to decide to share with a few trusted allies, any professional who may find it useful, or keep it to themselves.

\subsubsection{Build value-aligned communities}
Finally, the key to responsible practice revealed in our study and reinforced in existing scholarship is the role of community.
Like any other task in data science work, RDS practices rely on a network of helpful colleagues for explicit guidance but encouragement as well~\cite{pillai2026navigating,lin2026heres}. These networks take the form of public forums (e.g., Stack Overflow, Slack channels, in-person check-ins or workshops, accessible and secure channels like Signal)~\cite{deng2022exploring}. Previous work calls on RDS tools to integrate professional online communities to support responsible practice~\cite{deng2022exploring}.
Sometimes, as highlighted by \citet{widder2024epistemic}, there are no colleagues with similar values, and practitioners need to look outside of their institution to find like-minded professionals.

Therefore, we encourage tools to assess practitioners’ values and support the identification of professionals with shared values~\cite{widder2024power}. These relationships can set the stage for practitioners to swap notes, seek advice, collectively develop new strategies, and encourage sustained responsible practice through challenges~\cite{wong2023seeing}. 
However, value-based communities can lead to echo chambers, or in the worst cases, the development of communities bound by harmful values~\cite{lee2021viral}. Therefore, moderation mechanisms should be in place to guard against the devolution of professional RDS communities.
\subsection{Limitations}
Qualitative research and phenomenological analysis, especially, should not be seen as generalizable or representative of our research inquiry. There is a vast range of RDS techniques and professional data science sectors that were not represented in our study. Additionally, connectivity issues impacted the flow of conversation and participants' ability to directly interact with the intended workspace/board. P8's full recording and transcript were lost, and therefore their analysis was constructed based on reflection board content, interview notes, facilitator notes, and confirmatory research. P3 and P10's reflection board interactions were not captured in video recordings, and therefore board interaction analyses were gathered from transcripts and interview notes.

\section{Conclusion}
We examine \textit{``moral missions''} as phenomena of sustained responsible data science practice. In this paper, we present a semi-structured interview study with 15 responsible data scientists and AI practitioners to understand the moral decision-making procedures they use to articulate and actualize their convictions. We surface three interconnecting themes to capture our participants' approach to moral missions: 1) embodied introspection, 2) institutional circumvention, and 3) relationality. By unpacking the circumstances, moral analyses, and strategies of our participants, we also reveal pragmatic accounts of responsible GenAI decision-making. Finally, we contribute to the emerging scholarship of the (augmented) undercommons to call for subversive data science communities. In future work, we hope to co-design responsible data science tools with practitioners to realize virtual platforms and physical spaces for moral mission communities.
\section{Ethics Statement}
Given participants' professional commitments and obligations, we recognize the potential risks associated with discussing personal values, workplace practices, and organizational environments. Accordingly, participant privacy and maintaining confidentiality were prioritized throughout the research process with measures such as anonymization and data de-identification. Participants were offered opportunities to review recordings and transcripts, request the anonymization or removal of specific information, and withdraw from the study at any time with their data excluded. 
We also recognize that practices and strategies described in this work may have implications beyond the specific contexts within which they are described. While our focus is on those who identify as responsible data scientists and seek to enact values in service of social good, under different circumstances, approaches described in this work can be employed in ways that are exclusionary, harmful, or contrary to social good and public interest. Accordingly, we place significant attention on discussing implications of our findings and the directions they suggest for supporting RDS and RDS communities. 

\section{Positionality Statement}
The lead researcher is a Computer Science PhD student with experience in responsible data science as an intern, research assistant, and student. While this experience provided familiarity with responsible data science concepts and practices, she remains removed from the everyday experiences of many practicing data scientists. Given her research focus in data ethics, she remained attentive to the possibility of imposing theoretical frameworks onto participants' accounts and therefore sought to ground interview questions and follow-up discussions in participants' lived experiences. The lead researcher, along with several members of the research team, identifies with groups underrepresented in data science and computing. This informed a commitment to recruiting participants with diverse backgrounds and ensuring space for perspectives that may be less visible within data science research and practice.

The broader research team brings expertise, interests, and experience spanning computer science and information science, including responsible data science, data and AI ethics, human-computer interaction, data management and visualization, and AI policy. Team members bring varying levels of direct experience with data science practice, ranging from former non-professional practitioners and industry collaborators to researchers working in adjacent areas. We recognize that our disciplinary backgrounds, experiences, and commitments shape both our approach to research and the interpretations we develop. Throughout the research process, we sought to remain attentive to participants' perspectives and create space for them to describe their experiences in their own terms.

\section{Acknowledgments}
This work was supported in part by the National Science Foundation (Awards \# 2402718, \# 2514565, \# 2141506).

\bibliography{references}

\appendix
\section{Moral Missions: Supplementary Materials}
\subsection{Pre-Survey Questions} \label{sec:pre-survey}
After reviewing the study consent form, potential participants were invited to complete the following pre-survey. Optional questions are marked with an asterisk (*).

\begin{enumerate}
    \item Name. (short response)
    \item Pronouns.* (short response)
    \item What is your data science role? (short response)
    \item Years in data science? (numeric)
    \item How frequently do you complete the following data science tasks:
    \begin{enumerate}
        \item Categories: Ideation, Data Collection, Data Processing, Statistical Modeling, Testing \& Validation, Deployment, Monitoring, Communication (Data Visualization or Reporting)
        \item Responses: 5 (Often), 4, 3 (Rarely), 2, 1 (Never)
    \end{enumerate}
    \item In a 2019 keynote, Lise Getoor defines responsible data science as "efforts that address both the technical and societal issues in emerging data-driven technologies." How many years have you practiced responsible data science (numeric)
    \item Our study is primarily focused on responsible data science techniques. Some examples include fairness toolkits (Fairlearn), ethics checklists (Deon), documentation templates for data/model transparency (Datasheets for datasets), or problem ideation workbooks (Microsoft's Human-AI eXperience workbook). We are also very interested in ad hoc responsible data science techniques you have developed or encountered in your career (collection of real world examples). Please provide a short description of at least one responsible data science technique you have used or interested in using in your work. Feel free to return to this form to note more examples before the interview. (short response)
    \item Age.* (numeric)
    \item Gender.*
    \begin{enumerate}
        \item Non-binary
        \item Female
        \item Male
        \item Other, please specify
    \end{enumerate}
    \item Country of Residence (no abbreviation).* (short response)
    \item Ethnicity (check all that apply).*
    \begin{enumerate}
        \item African American, African, and/or Afro-Caribbean
        \item Hispanic or Latino
        \item Asian or Pacific Islander
        \item Middle Eastern or North African
        \item Native American or Alaska Native
        \item European or White
    \end{enumerate}
    \item Highest Degree Earned.*
    \begin{enumerate}
        \item Highschool
        \item Certificate
        \item Associates
        \item Bachelors
        \item Masters
        \item Doctorate
    \end{enumerate}
    \item Income range (in USD).*
    \begin{enumerate}
        \item Less than \$30,000
        \item \$30,000 - \$70,000
        \item \$70,000 - \$100,000
        \item Above \$100,000
    \end{enumerate}
    \item After reviewing the consent form and the pre-survey questions, do you consent to this study? If so, please type your full name below. (short response)
    \item Do you have any accessibility needs for the virtual interview? (examples: closed captioning, interpreter/translator, access to questions before interview, etc.).* (long response)
\end{enumerate}

\subsection{Semi-structured Interview Questions} \label{sec:interview}
Along with the moral missions workspace, the following interview questions guided the semi-structured interview with participants. The sub-questions serve as potential probe or follow up questions.
\begin{enumerate}
    \item Can you tell me a little about your work?
    \begin{enumerate}
        \item In the pre-survey you noted you have been a RDS for [timeframe], how would you define being a responsible data scientist?
        \item What specific technology do you build, work on or work with?
        \item Do you work with a team of other data scientists, an interdisciplinary team, or primarily by yourself?
        \item How did you gain your data science skills? Do you have a background in fields beyond data science?
        \item Who do you build technology for?
        \item Who are the specific populations or communities represented in the data you work with?
    \end{enumerate}
    \item What values are most important to you in your responsible data science work? Please choose 3 - 5 values from this collection. If there is a value you care about that's not in this list, feel free to use the template card to add it to the board and provide your definition. Feel free to think aloud as you select them.
    \begin{enumerate}
        \item Why did you select each value? Feel free to adjust the definition to your understanding.
        \item What experiences, even beyond your profession, reinforced these values as important?
        \item Are these values informed by a larger mission or philosophy you have about social progress?
    \end{enumerate}
    \item Now, in the pre-survey, you indicated [responsible data science technique] is a technique you have used in your work. For the rest of the interview, we will discuss your process and considerations for using this technique. Would you like to discuss this technique or select another?
    \item From the values you selected earlier, which values are addressed by this technique?
    \begin{enumerate}
        \item Can rank the relevance/importance of these values for the technique?
    \end{enumerate}
    \item How did you find out about this technique? 
    \begin{enumerate}
        \item If someone shared this technique with you, how did they explain it and what were the circumstances in which they shared it?
        \item If you found it through online resources, what are these resources?
        \item If you read books or academic papers, how closely/how much did you review/read the content?
        \item Or did you develop or brainstorm this idea on your own? If so, please describe that process?
        \item If you used other means, describe how you used these resources?
    \end{enumerate}
    \item How have you adapted this technique across circumstances?
    \begin{enumerate}
        \item Can you describe a project or role you really enjoy working on that uses this technique?
        \item Can you describe a project or role you struggled with or didn’t enjoy working on that uses this technique?
    \end{enumerate}
    \item What criteria do you have for when to use the technique?
    \begin{enumerate}
        \item What technical, ethical, or social aspects did you consider?
        \item How did you realize an adjustment had to be made in your implementation compared to other experiences?
        \item Did you consult anyone to work through the solution?
    \end{enumerate}
    \item How has this technique developed over time due to your personal understanding?
    \begin{enumerate}
        \item How do you self-evaluate if your implementation addresses the social considerations you intended?
        \item Is there documentation or formalization of this process?
    \end{enumerate}
    \item Let’s review your board. Do you think the techniques we discussed today actualize the values you selected? 
    \begin{enumerate}
        \item If so, why do you think so? Please feel free to draw connections between the values and the techniques as you see fit.
        \item If not, how could you imagine further adjusting your techniques to serve these goals?
        \item Additionally, would you like to revise or add nuance to the values you provided?
        \item For the communities impacted by your work, are their values reflected in yours? It’s ok if not but if you would like to amend or add values you can.
    \end{enumerate}
    \item What external factors make it easier or harder to achieve these values in your data science work?
    \item Are there any other aspects of your process for using responsible data science techniques that we did not cover?
    \item Are there techniques you are interested in using but have yet to?
    \item What do you think the overall responsible data science community is missing in their practices or techniques?
\end{enumerate}

\subsection{Interview Workspace}
\begin{figure}[ht!]
    \centering
        \includegraphics[width=0.75\linewidth]{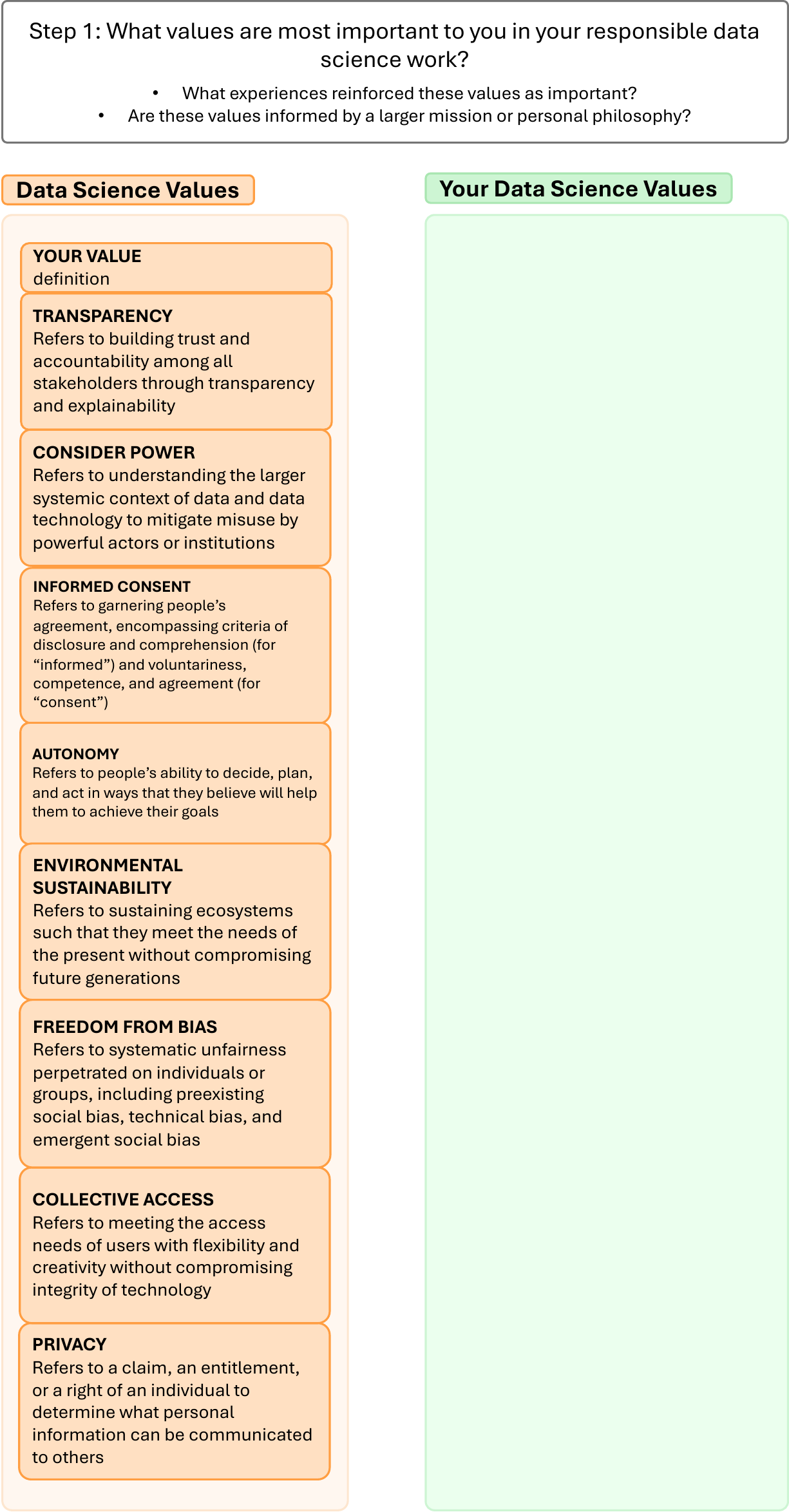}
    \caption{A diagram showing the layout of value cards used in our study. Participants could also write their own custom value cards.}
    \label{fig:step1-valuecards}
\end{figure}
\subsubsection{Value Cards}
Once the FigJam workspace was introduced to the interview participant, we presented a collection of value cards inspired by \citet{friedman2012envisioning, shen2021value}. The value cards were prepared by reviewing a diverse array of manuscripts that present a framework of principles or a collection of values relevant to RDS practice. A subset of representative values across the papers were selected and summarized to create each card depicted below. 

We instructed the participants to select three to five cards so participants can discuss the breadth of their data ethics perspective within the session time. Interview facilitators stressed that the presented value cards are informed by prior literature but not exhaustive. Therefore, a template value card was placed first to encourage participants to define data ethics values. The selected value cards were referenced throughout the rest of the interview.
\subsubsection{RDS Technique Cards}
With the values cards selected, we turned the interview to responsible data science practices. In the pre-survey, participants were asked to provide a responsible data science technique they would feel comfortable discussing in the interview. We paraphrased the technique detailed in the presurvey response and participants had the choice to discuss the technique or choose a new one. To inspire on the spot technique selection, we prepared techniques cards to present categories of common RDS techniques~\cite{kuehnert2025who}. 

After selecting a technique, participants were asked to name and provide a brief definition of the technique. Then, as depicted in \pageref{fig:step2-techniquecards}, the interviewer lead the participant through four prompts (see questions five through eight in \pageref{sec:interview}). First, participants were asked to select personal value cards (from the previous step) that were relevant to their technique. Participants were also encouraged to rank or relate the value cards to each other. Participants were encouraged to use sticky notes, external links, images, and more. 

\subsubsection{Value-laden Reflection}
In the final set of questions (see nine through 13 in \pageref{sec:interview}), we ask participants recontextualize their values based on their technical experience. These questions also aimed to create a space for participants to discuss contexts out of their control that impact their values and responsible practice. This often led participants to imagine what larger systemic change is needed to improve their responsible technical practice.
\begin{figure}[ht!]
    \begin{subfigure}{0.5\textwidth}
        \includegraphics[width=1\linewidth]{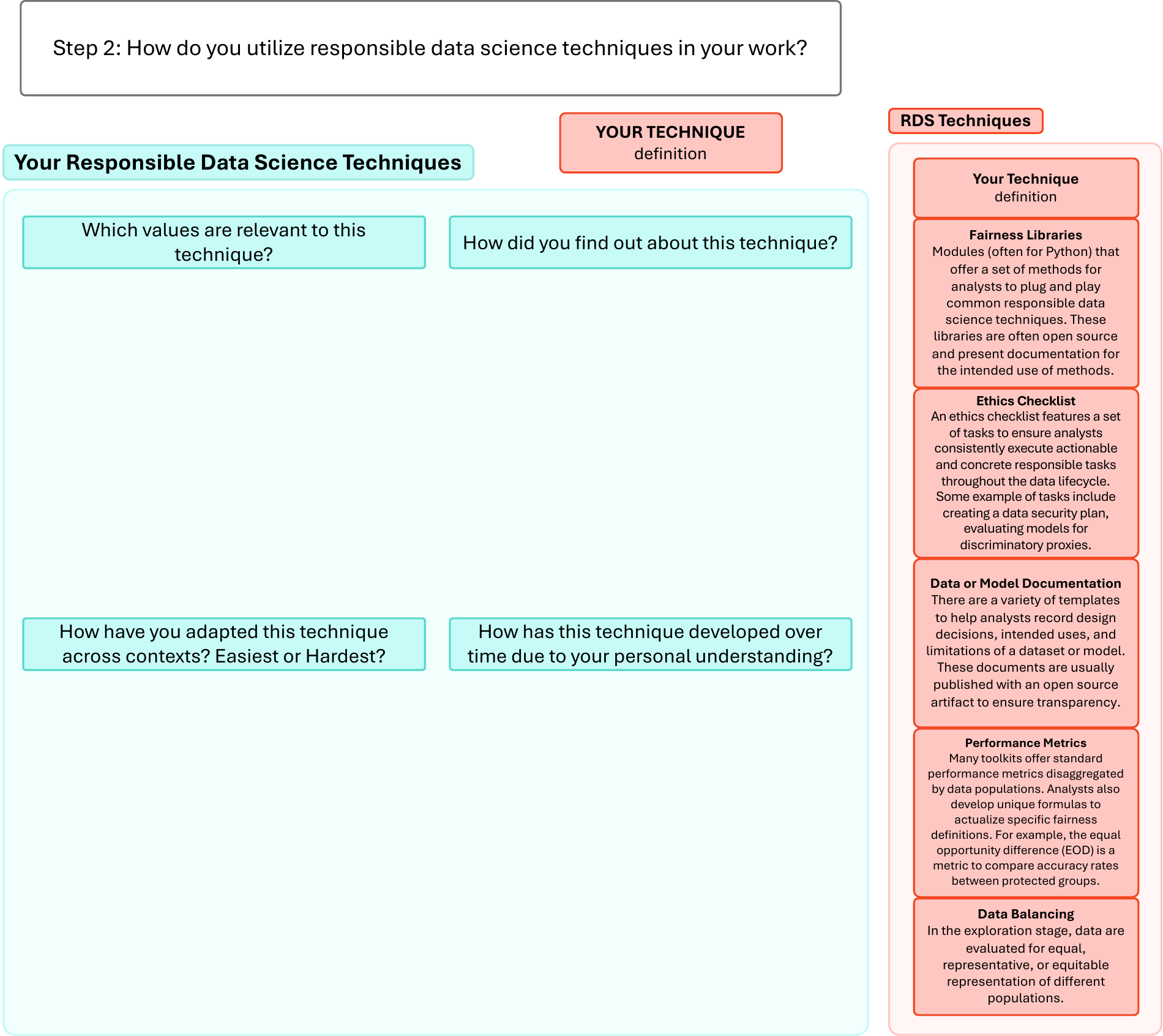}
    \caption{A diagram showing the layout of technique cards used in our study. Participants could also write their own custom technique cards.}
    \label{fig:step2-techniquecards}
    \end{subfigure}
    \begin{subfigure}{0.5\textwidth}
        \includegraphics[width=1\linewidth]{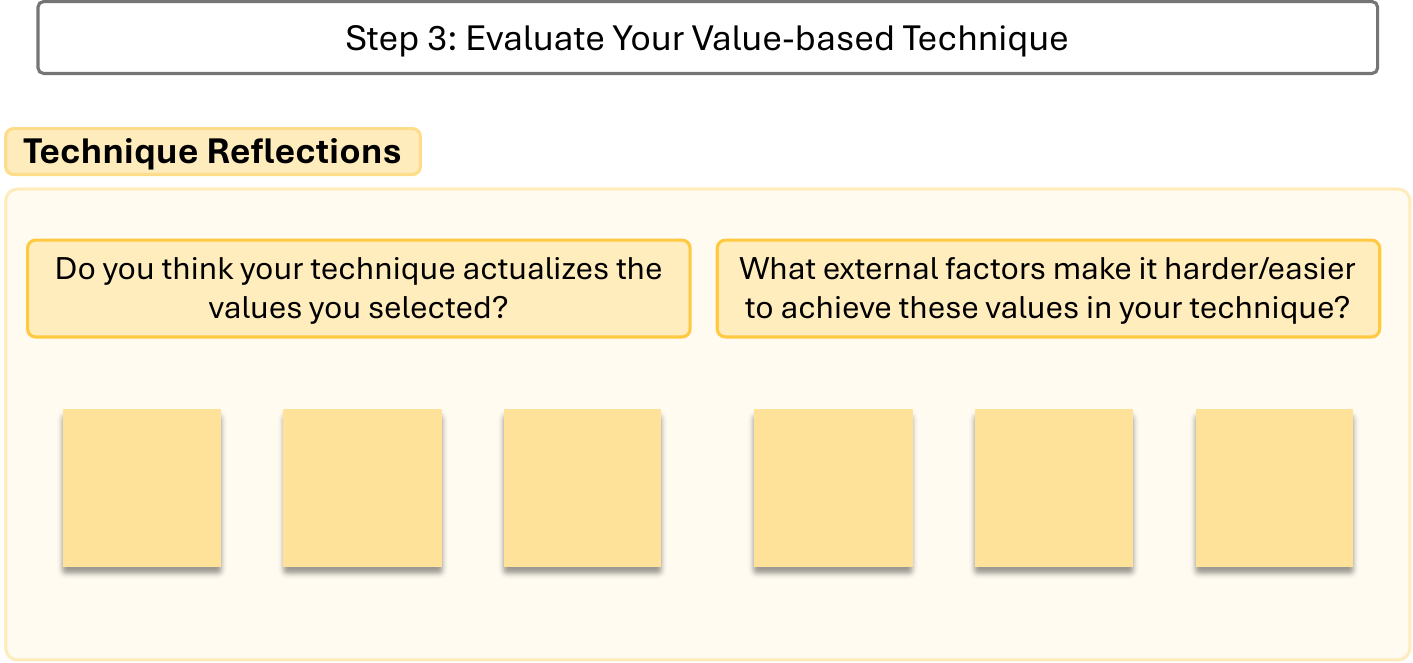}
    \caption{A diagram showing the layout of the reflection portion of our study.}
    \label{fig:step3-reflection}
    \end{subfigure}
\end{figure}

\begin{figure}[ht!]
    \begin{subfigure}{0.5\textwidth}
    \includegraphics[width=1\linewidth]{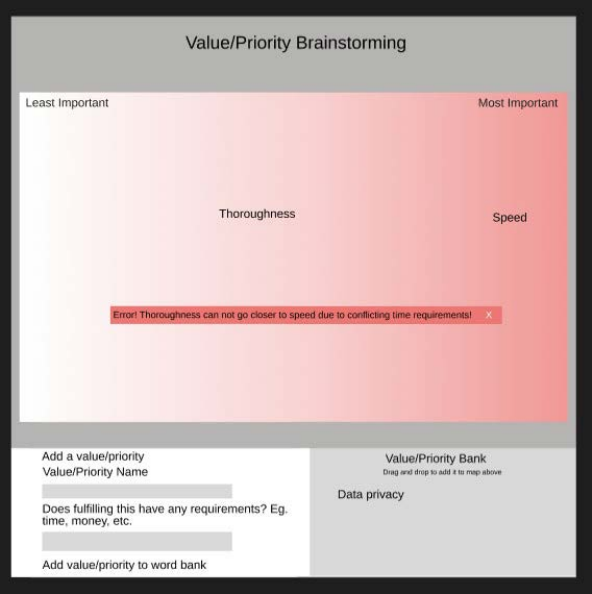}
    \caption{\textbf{Develop moral agency} Inspired by P12, a user can visually reason about the relationships and priorities or their values. A user can add value with details of what it requires. Then, the submitted words can be placed on a two-dimensional spectrum of least to most important. A warning flashes when values with conflicting requirements are placed too close together.}
    \end{subfigure}
    \begin{subfigure}{0.5\textwidth}
    \includegraphics[width=0.9\linewidth]{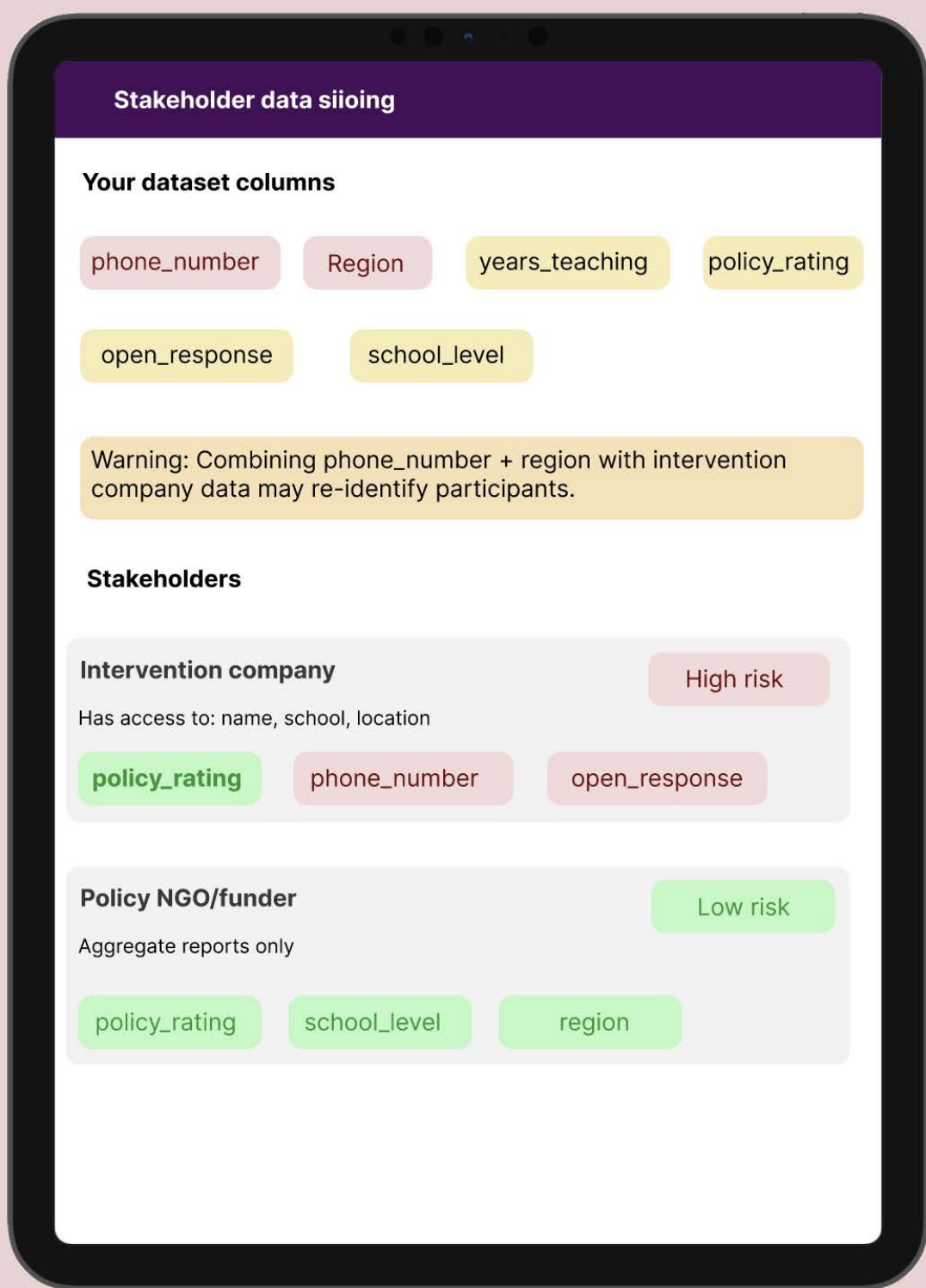}
    \caption{\textbf{Maintain sociotechnical strategies} Inspired by P8, a user determines a data siloing strategy that considers data columns to exclude, warnings, and stakeholder data access profiles. These widgets reflect the key decision-making components of P8's data siloing considerations and records the strategy for dissemination with team members or reflection.}
    \end{subfigure}
\end{figure}
\begin{figure}[ht!]
    \centering
    \includegraphics[width=1\textwidth]{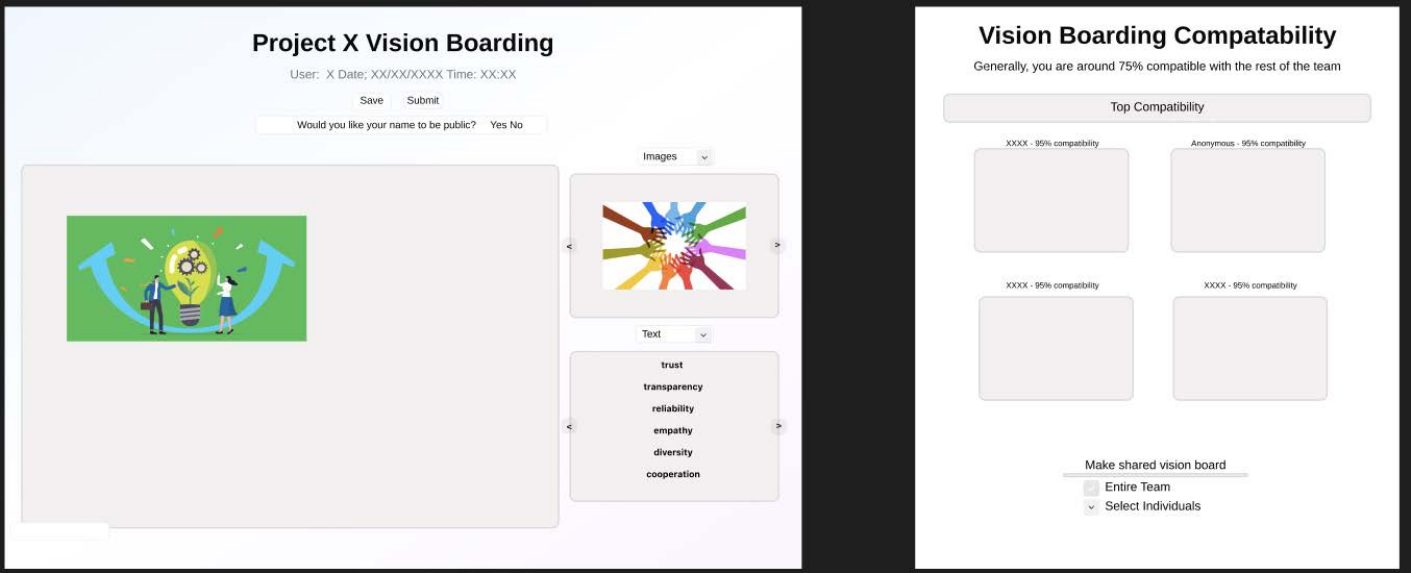}
    \caption{\textbf{Build value-aligned communities} Inspired by P2's account, a user creates a multimedia vision board to depict value-laden convictions that shape their responsible data science practice. Then, based on keywords used in the vision board, this imagined system matches other professionals with similar values. The user can then choose to create joint vision boards with their project team or select individuals they are compatible with.}
\end{figure}
\subsection{Design Implications Rapid Prototyping Exemplars} \label{sec:example}
We present three curated examples to demonstrate possible directions for each design implications.

\end{document}